# Dual-comb spectroscopy using free-running mechanical sharing dual-comb fiber lasers

**Haochen Tian,**[1,2] **Runmin Li,**[1] **Takeru Endo,**[1] **Takashi Kato,**[1,4] **Akifumi Asahara,**[1] **Lukasz A. Sterczewski,**[3] **and Kaoru Minoshima**[1,*]

[1]Graduate School of Informatics and Engineering, The University of Electro-Communications, 1-5-1 Chofugaoka, Chofu, Tokyo 182-8585, Japan
[2]JSPS Postdoctoral Fellowships for Research in Japan, 1-5-1 Chofugaoka, Chofu, Tokyo 182-8585, Japan
[3]Faculty of Electronics, Photonics and Microsystems, Wroclaw University of Science and Technology, Wybrzeże Wyspiańskiego 27, 50-370 Wrocław, Poland
[4]PRESTO, JST, 1-5-1 Chofugaoka, Chofu, Tokyo, 182-8585, Japan

**Abstract**. We demonstrate balanced-detection dual-comb spectroscopy (DCS) using two free-running mechanical sharing dual-comb fiber lasers assisted by an all-computational digital phase correction algorithm. The mutual coherence between the combs allows us perform mode-resolved spectroscopy of gaseous hydrogen cyanide by digitally compensating residual timing and offset frequency fluctuations of the dual-comb signal. Setting the repetition rate difference between the combs to 500 Hz (1.5 kHz) yields more than 2000 resolved radio frequency comb lines after phase correction in a 3-dB bandwidth centered at 1560 nm of wavelength. Through coadding the corrected interferograms (IGMs), we obtain a single time-domain trace with a SNR of 6378 (13960) and 12.64 (13.77) bits of dynamic range in 1 second of averaging. The spectral SNR of the coadded trace reaches 529 (585), corresponding to a figure of merit of SNR of $1.3 \times 10^6$ ($1.4 \times 10^6$). The measured absorption spectrum of hydrogen cyanide agrees well with the HITRAN database.

**\***Corresponding Author, E-mail: k.minoshima@uec.ac.jp.

A pair of mutually coherent optical frequency combs (OFCs) with offset repetition rates has emerged as a powerful tool in many spectroscopic applications like multidimensional spectroscopy [1,2], distance measurement [3], holography [4] or hyperspectral imaging [5]. The dual-comb technique that converts information from the optical to electrical domain via multi-heterodyne beating [2] was initially confined to fully-stabilized, phase-locked sources. This has changed recently, when dual-comb (or even multi-comb) emission from a free-running, single laser cavity has proven to be an effective way of obtaining moderate-coherence sources with relative stability sufficient for practical purposes. Wavelength-multiplexing [6,7], polarization-multiplexing [8], circulation-direction-multiplexing [9], or cavity-space-multiplexing [10] have been employed for this free-running dual-comb generation scheme [11]. Unfortunately, most of these laser configurations still require specific cavity design and non-trivial cavity alignment with restrictions on the tuning range of the repetition rate difference. Recently, we proposed mechanical sharing of laser cavities as a method to obtain high mutual coherence between the two combs that offers greater agility in obtainable repetition rates and repetition rate difference [12]. Owing to effective common-mode noise suppression, the relative 3-dB linewidth between the two combs was measured to be ~ 50 Hz in 38 ms.

Although free-running dual-comb lasers have already been employed for comb-line-resolved absorption spectroscopy of gaseous samples [13], the relative repetition rate and offset frequency in such cavities still exhibits long-term drifts owing to residual non-common-mode noise induced by environmental fluctuations. This degrades the long-term mutual coherence and results in phase fluctuations of the measured electrical signal (interferogram, IGM), which in turn prevent coherent averaging to achieve a high signal-to-noise ratio (SNR). To address this issue and effectively suppress phase noise, various data post-processing and digital phase correction techniques can be utilized. One approach relies on introducing free-running CW lasers as intermediate oscillators. The IGMs' phase fluctuations in this scenario are corrected using signals extracted from the beats between the combs and CW lasers [14-16]. Hybrid techniques utilizing slow phase-locked loops (PLL) assisted by digital extraction of correction signals have also been proposed [17]. Most importantly, it is possible to extract correction signals solely from the measured IGM, i.e. without any additional phase locking, photodetectors (PDs) or auxiliary lasers [18-22]. These all-computational approaches are essential for less mature spectroscopic regions like the mid-infrared or terahertz, where the availability of optoelectronic components is scarce. Unfortunately, this convenience has its price. Owing to the absence of an external reference, the short-term frequency stability between the combs must be higher than $\Delta f_{\text{rep}}$ on a $1/\Delta f_{\text{rep}}$ timescale, where $\Delta f_{\text{rep}}$ is the repetition rate difference. If this condition is violated, comb line frequencies change by more than half the line spacing between two consecutive IGM frames. Since this effect cannot be easily identified due to the violation of the Nyquist criterion, this level of relative stability is a prerequisite for mode-resolved spectroscopy using all-computational algorithms. The mechanical sharing scheme discussed here complies with this requirement, which gets rid of the additional optical intermedia, photodetectors, and digitization channels [16].

In addition to phase noise, amplitude noise also lowers the obtainable precision. The SNR of the detected IGMs increases proportionally with optical power except for the saturation regime, when the PD starts to generate spurious signals due to excessive optoelectronic nonlinearities. There have been several methods to avoid this issue. In the first, the optical power limitation can be overcome by stretching the incident optical pulses with a fiber Bragg grating before detection. This way, one reduces the pulse peak power being responsible for the highly nonlinear response. Via digital deconvolution, it is then possible to remove the IGM chirp in post-processing [14]. In the second approach, the PD nonlinearity can be digitally corrected by estimating the PD's electrical response from spurious frequency components like harmonics or intermodulation products [23]. Any of these methods needs to be employed when the PD operates close or beyond its limits. Here we solve this issue by simply improving the photodetector performance.

Overall, this work aims to establish a simple and practical scheme for sensitive free-running dual-comb spectroscopy (DCS). To obtain DCS IGMs, we beat two free-running all-polarization-maintaining mechanical sharing combs on a home-made balanced photodetector. Because, its noise floor is much lower than that of two commercially-available ones, we obtain an SNR improvement critical for enhancing the absorption sensitivity of DCS measurements. Picometer-resolution and phase noise suppression are obtained by means of computational phase correction. From the measured data we conclude that the short-term performance of our DCS system is close to state-of-the-art fully-phase-locked systems. The combination of mechanical sharing combs and all-computational digital phase correction provides a simple and practical tool for picometer-level-resolution DCS with prolonged averaging capabilities and suitability for future translation to other spectral regions.

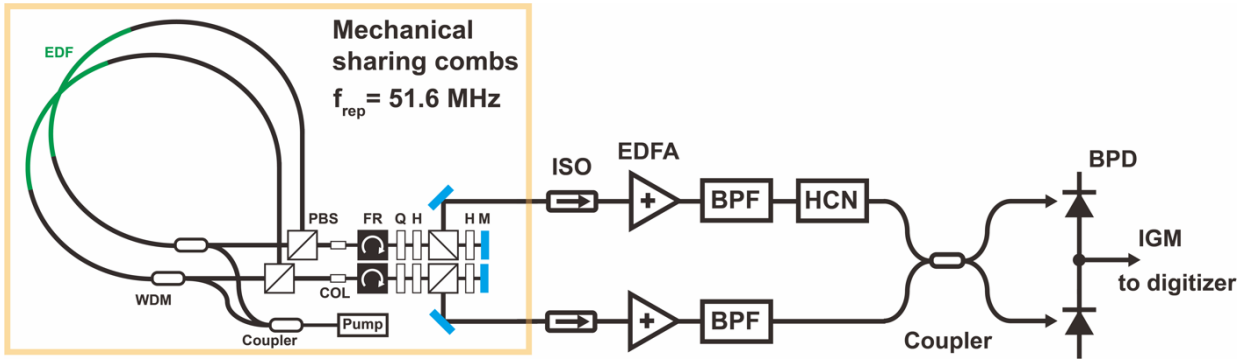

**Fig. 1.** Experimental setup of the dual-comb spectroscopy using mechanical sharing combs. EDF: Er-doped fiber; WDM: wavelength division multiplexer; PBS: polarizing beam splitter; FR: faraday rotator; Q: quarter waveplate; H: half waveplate; M: mirror; ISO: isolator; EDFA: Er-doped fiber amplifier; BPF: bandpass filter; HCN: hydrogen cyanide gas cell; BPD: balanced photodetector.

The experimental setup of our dual-comb spectroscopy system is shown in Fig. 1. The outputs of the two mechanical sharing combs are guided through two fiber isolators, two home-made fiber amplifiers, two tunable optical bandpass filters, and are finally combined by a 2×2 fiber coupler. The repetition rates of the two combs are set to ~ 51.6 MHz [12]. For gas absorption measurements, we inserted a hydrogen cyanide ($H^{13}CN$) gas cell after the bandpass filter into one arm of the DCS spectrometer. To realize balanced detection of DCS IGMs instead of a single PD, the two ports of the home-made BPD are simultaneously illuminated with 50 µW of optical power each. We found that in the linear response regime, the performance of our home-made BPD is higher than that of Newport (1817) and Thorlabs (PDB415C) BPDs. For comparative purposes, we use a microwave spectrum analyzer to obtain their spectral SNR that is defined as the difference between the spectral peak and the detector noise floor, amounting here to 46 dB, 35 dB and 36 dB for the home-made BPD, Thorlabs BPD, and New focus BPD case, respectively. Also in the time-domain (IGM peak versus noise standard deviation away from the center burst), we obtain a single-shot SNR of 403 (52 dB), 229 (47 dB) and 234 (47 dB) for the home-made, Thorlabs, and New focus BPD, respectively, accessed with a 14-bit digitizer without phase correction. The lower value of the spectral SNR can partially be attributed to weak spectral aliasing, because the DCS spectral bandwidth exceeds the first Nyquist zone when one includes components lying −50 dB relative to the peak (and become relevant here). Nevertheless, in both domains our home-made BPD outperforms the tested commercial ones in terms of obtainable SNR. Detailed information and experimental results can be found in Supplementary Material.

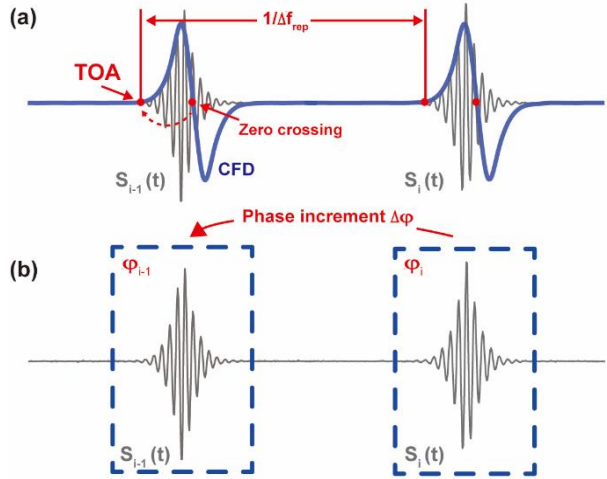

**Fig. 2.** Principle of the determination of (a) TOA and (b) offset phase increment.

The IGMs resulting from periodic interference between the two mechanical sharing combs have shot-to-shot fluctuations in carrier frequency and timing. These arise from relative instabilities of the combs' repetition rates ($\Delta f_{\text{rep}}$) and carrier-envelope offset frequencies ($\Delta f_{\text{ceo}}$). Directly averaging such phase-noisy IGMs would smear out the analyte's fine absorption features (both in amplitude and phase). Nevertheless, the IGM digital phase correction algorithm used previously for polarization-multiplexed DCS can mitigate this issue [19] given that the laser used here meets the correctability criterion (relative frequency stability of ~ 50 Hz in 38 ms, upper limit estimation).

The first correction step is the precise determination of the time of arrival (TOA) of every IGM center burst. The imaginary part of the sampled IGM, $S_{im}(t)$ is generated through the Hilbert transform to next extract the IGM envelope, $S_{env}(t)$:

$$S_{env}(t) = |S(t) + i \cdot S_{im}(t)| \quad (1)$$

where $S(t)$ is the sampled IGM. To precisely determine the timing of IGM frames, we use a constant fraction discriminator (CFD). The extracted envelope is delayed and subtracted from the original to generate an S-shaped CFD signal, shown as the blue curve in Fig. 2 (a). The 150[th] data point ahead of the zero-crossing point of the CFD signal is regarded as the IGM's TOA, as shown in Fig. 2 (a). Similar to the idea of balanced optical cross-correlation technique in timing jitter measurements of optical pulses [24,25], the CFD block utilizes subtraction to lower the sensitivity to IGM intensity fluctuations, which corrupt the estimation of TOA. Using TOAs of all IGM center bursts, we obtain durations of consecutive IGM frames. Next, we resampled all of them to the same duration by linear interpolation. This way, repetition rate fluctuations of the IGM are suppressed assuming linear phase drift between each IGM frame.

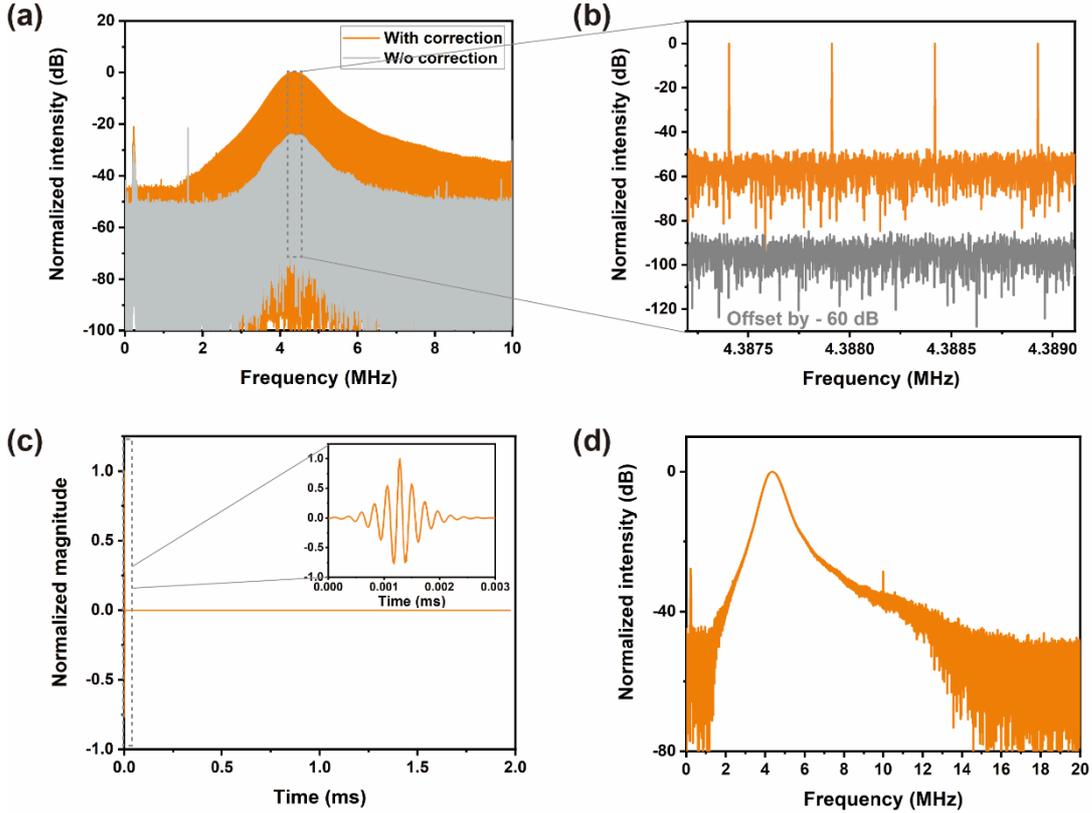

**Fig. 3.** Digital correction results. (a) Spectra of the IGM stream without (gray curve) and with (orange curve) digital correction. (b) Zoomed view of comb modes. (c) Time-domain trace of coherently averaged ~500 IGMs in 1 s. (d) Spectrum of (c).

The second correction step is to compensate relative offset frequency fluctuations. Data points around the center burst are extracted, as shown in Fig. 2 (b). The average phase increment between each pair of IGMs is calculated by [26]:

$$\Delta\varphi_i = \text{mean}\left\{\text{angle}\left[S_i(t) \times S_{i-1}^*(t)\right]\right\} \tag{2}$$

where $S_i(t)$ is the current IGM, $S_{i-1}(t)$ is the previous IGM, the asterisk denotes complex conjugate. The current IGM is next multiplied by $\exp\left\{i \cdot \left[\Delta\varphi_i + \left(\sum_{k=1}^{i-1}\Delta\varphi_k\right)\text{mod}\,2\pi\right]\right\}$, leading to the cancellation of offset frequency fluctuations in every IGM pair. In this way, a carrier-envelope-offset-free stream of IGMs is obtained. The abovementioned correction steps effectively restore the discrete comb nature of the down-converted dual-comb spectrum.

To verify the efficacy of phase noise suppression in our system, we apply digital correction to IGMs first without the analyte. The repetition rate difference between the two combs is set to ~500 Hz, which yields ~500 IGM frames per second, digitized at a 50 MHz sampling rate. After applying repetition rate correction, IGM frame durations almost do not show any fluctuations, as shown in Fig. S3 (a) in the supplementary material. This is also confirmed in the digital difference frequency generation (DDFG) signal [19]. Its spectrum that yields the fundamental and harmonics of the IGMs' repetition rate has narrowed linewidths, while the lines free of noise pedestals. Following timing correction, offset frequency compensation is applied to the repetition-rate-corrected IGM frames.

The spectrum of the IGM sequence (containing ~500 IGM frames) before the phase correction is shown as gray curve in Fig. 3 (a). Phase instabilities in the IGMs smear out the discrete character of RF comb lines. In contrast the corrected RF spectrum (orange trace in Fig. 3 (a)) calculated from stitched corrected IGM frames has a SNR that is improved by ~23 dB. Analyzing a zoomed portion of the RF spectrum shown in Fig. 3(b), it is also clear that the phase and timing correction restores the initially corrupted discrete shape of the RF spectrum and suppresses spurious non-comb lines. The RF linewidth of the mode-resolved RF spectrum is ~1 Hz, which is limited by the acquisition time, while the line-to-floor ratio (LFR) reaches ~50 dB [27]. Because the ~500 corrected IGM frames are all in phase, they can be averaged to produce a single time-domain trace with a high dynamic range. The result of this operation is shown in Fig. 3 (c). The averaged IGM has a time-domain SNR of 6378 in 1 s, while before the correction, the single-shot SNR is only 403, (see Table S1 in the supplementary material). Unlike other DCS systems that generate two combs from a single laser cavity, the flexible tunability of $\Delta f_{rep}$ between the two mechanical sharing lasers allow us to investigate the performance of the DCS system with different $\Delta f_{rep}$. The coadded 1500 IGMs after correction with 1.5-kHz $\Delta f_{rep}$ are shown in Fig. 4 (a). In this case, the SNR reaches 13960. Intuitively, the increased number of averaged IGMs within the same acquisition time leads to an SNR improvement.

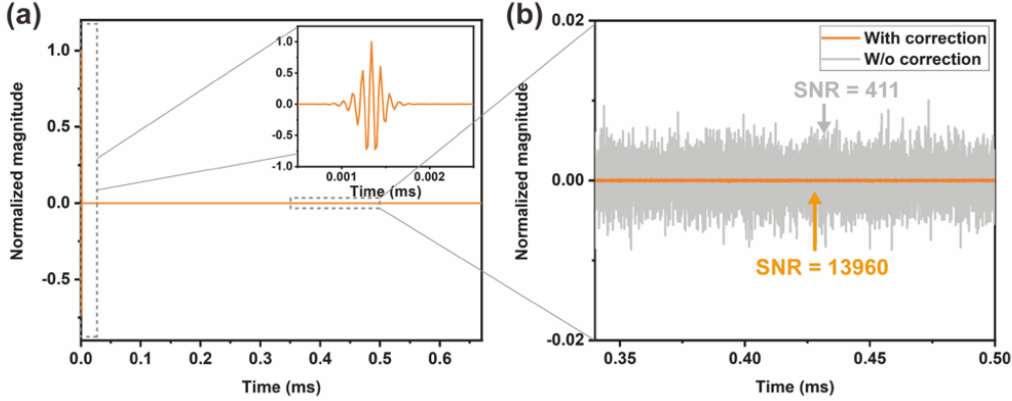

**Fig. 4.** (a) Coadded time-domain trace of ~ 1500 IGMs in 1 s when $\Delta f_{rep}$ is 1.5 kHz. (b) Zoomed view of the trace away from the center burst.

It is imperative to compare our system with fully phase locked systems. The dynamic range of the time domain trace in units of bits is calculated through $\log_2(SNR_{tr})$, where $SNR_{tr}$ denotes the SNR of the time domain trace [28]. When $\Delta f_{rep}$ is ~500 Hz (1.5 kHz), the dynamic range of the coadded IGM in our system yields 12.64 (13.77) bits over 1 s. This value drops to 10.85 (11.97) bits when taking into account the variance of a uniformly distributed noise due to quantization [29]. Therefore our system is comparative (> 90% correction efficacy) to phase-locked systems in terms of dynamic range. The spectrum of the coherently averaged IGM is shown in Fig. 3 (d). We can also compare the spectral SNR from the standard deviation of the baseline in addition to the dynamic range. We obtain a value of 529 (585) over 1 s in a 123 GHz span around a 192.17 THz central optical frequency, corresponding to a SNR figure of merit (FoM) of $1.3\times10^6$ ($1.4\times10^6$). This reaches typical levels as summarized in the review paper [2]. Given the significantly simplified experimental setup, the minor deterioration in performance is acceptable.

To assess the spectroscopic performance of the presented DCS system, a hydrogen cyanide gaseous cell (25-torr, 22 cm, room temperature, Wavelength References) is inserted after the bandpass filter into one arm of the spectrometer. It is rational to use free-running combs with 51 MHz repetition rate here because the absorption lines of the sample under test are much wider than the 51 MHz comb's resolution. We sample and correct ~1500 IGMs when $\Delta f_{rep}$ is set to be around 1.5 kHz. The corrected IGMs are co-added and Fourier-transformed to obtain the power spectrum, as shown in orange curve of Fig. 5 (a). Absorption features located at 1558.03 nm (line *P*20), 1558.91 nm (line *P*21), 1559.81 nm (line *P*22) and 1560.71 nm (line *P*23) can be observed in the DCS spectrum.

For spectroscopic performance evaluation, we focus on line *P*22 line and compare it with the HITRAN database. To calculate the optical transmission, we fit the spectrum centered at 1558.91 nm across ~ 100 GHz window with a $5^{th}$-order polynomial to serve as a synthetic zero-gas measurement [30]. Using an empty gaseous cell as sample is alternative to obtain true zero-gas measurement. Because only one comb interrogates the sample, we retrieve the transmission from the ratio of the power spectrum to the synthetic zero-gas spectrum, as shown with red scatters in Fig. 5 (b). The figure additionally includes a Voigt fit plotted with solid line. The transmission modeled using the HITRAN database and that retrieved from the DCS measurement are plotted in mirrored panels for comparative purposes. Overall they show good mutual agreement except for periodic fringes resulting from imperfections of optical coatings inside the absorption cell that form optical etalons. The fitted Voigt full-width at half-maximum (FWHM) of 8.6 pm indicates that the molecular transition is Doppler-limited and can be probed with picometer-level resolution.

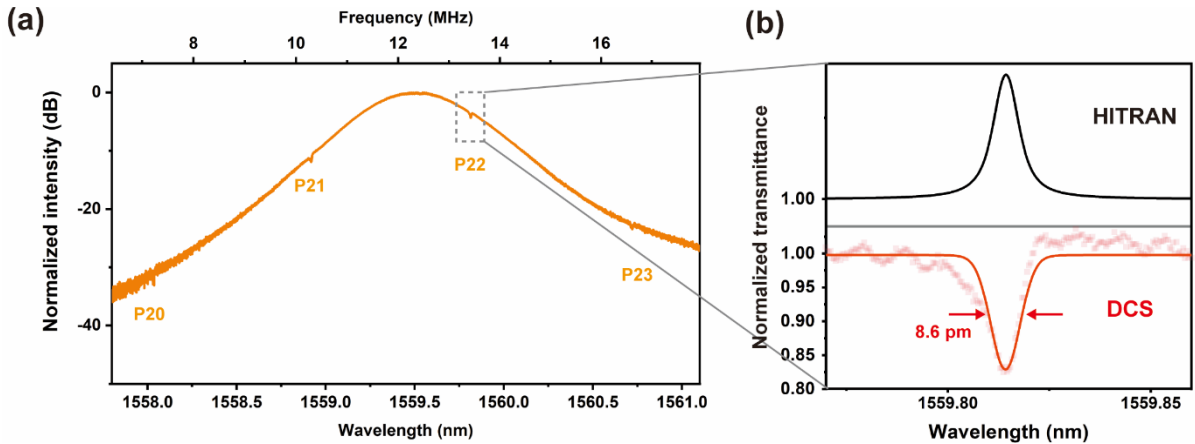

**Fig. 5.** Absorption profile measurement of a hydrogen cyanide gaseous cell. (a) The spectrum of the coadded IGM. (b) The normalized transmittance of *P*22 line. The measured absorbance from DCS measurement, its Voigt fitting and the absorbance from HITRAN database are represented as red scatters, red curve and black curve, respectively.

We would like to note here that in the same measurement performed with $\Delta f_{rep}$ of ~500 Hz, we observe absorption linewidth broadening to 15 pm in 1 s averaging time (~500 IGMs). This proves that higher IGM update rate is beneficial for avoiding excessive aliasing of noise above $\Delta f_{rep}$. It lowers the probability of phase correction failure and brings the performance of our free-running DCS system approach closer to phase-locked dual-comb lasers [31]. Thus ~ kHz $\Delta f_{rep}$ is preferable for pm-level spectroscopy measurements when free-running dual-comb lasers with ~ 100 MHz comb spacing are used [13,19].

In summary, we realize balanced-detection dual-comb spectroscopy using two free-running mechanical sharing fiber combs assisted by an all-computational phase correction algorithm. After two steps of correction, the discrete characteristics of RF comb lines can be restored. 2387 comb modes spanning a 123-GHz range (3-dB bandwidth) centered at 192.17 THz optical frequency are resolved. When the $\Delta f_{rep}$ is ~ 500 Hz (~ 1.5 kHz), ~500 (~ 1500) corrected IGMs acquired over 1 s are coadded to obtain a single time-domain trace with an enhanced SNR of 6378 (13960) and a dynamic range 12.64 (13.77) bits. The spectral SNR of the coadded IGM is 529 (585), corresponding to a SNR FoM of $1.3 \times 10^6$ ($1.4 \times 10^6$). In the down-converted DCS spectrum we observe four absorption lines of gaseous $H^{13}CN$ (lines $P$20 – $P$23). A representative fit to transmittance data around line $P$22 shows good agreement with the HITRAN database.

Owing to the combination of polarization-maintaining mechanical sharing combs and all-computational correction algorithm, our experimental setup requires no bulky phase locking circuitry or external optical references while permitting extended-timescale coherent averaging. Instead of the currently limited 3 nm spectroscopic coverage, >10 nm broadband measurements are possible when a programmable optical tunable band pass filter is implemented and then the down sampled RF spectra obtained from distinct wavelengths are stitched [13]. Compared with reported literatures on DCS measurements using dual combs generated from a single cavity in the dual-wavelength scheme [13, 32], our system does not require spectral broadening to ensure mutual spectral overlap between the combs and allows for a more flexible selection in $\Delta f_{rep}$. The fact that two independent oscillators with a shared mechanical environment are suitable for mode-resolved DCS is of utmost importance because this configuration is virtually free of soliton-soliton interactions which would become a source of increased intensity and phase noise in single-cavity configurations. Here, we obtain clean IGM trances free of any spurious components that plague other systems. We believe that the experimental configuration presented here promises easy translation to other spectroscopically relevant regions like the mid-infrared or even terahertz in the future.

*Supplementary material*

See supplementary material for detailed scheme of the home-made BPD, its performance comparison with commercial BPDs, etc.

*Funding*

This work was supported by Japan Society for the Promotion of Science (JP21H05014); Horizon 2020 Framework Programme (H2020) (101027721).

*Acknowledgments*

H. Tian and R. Li acknowledge funding from Japan Society for the Promotion of Science (JSPS). L. A. Sterczewski acknowledges funding from the European Union's Horizon 2020 research and innovation programme under the Marie Skłodowska-Curie grant agreement No 101027721. The authors thank M. Endo in University of Tokyo for useful discussion.

*Disclosures*

The authors declare that there are no conflicts of interest related to this article.

*Data availability*

The data that support the findings of this study are available from the corresponding author upon reasonable request.